\documentclass{IEEEoj}
\usepackage{cite}
\usepackage{amsmath,amssymb,amsfonts}
\usepackage{algorithmic}
\usepackage{graphicx,color}
\usepackage{textcomp}
\def\BibTeX{{\rm B\kern-.05em{\sc i\kern-.025em b}\kern-.08em
    T\kern-.1667em\lower.7ex\hbox{E}\kern-.125emX}}
\AtBeginDocument{\definecolor{ojcolor}{cmyk}{0.93,0.59,0.15,0.02}}

\begin{document}
\receiveddate{XX Month, XXXX}
\reviseddate{XX Month, XXXX}
\accepteddate{XX Month, XXXX}
 \publisheddate{XX Month, XXXX}
 \currentdate{XX Month, XXXX}

\title{LapGSR: Laplacian Reconstructive Network for Guided Thermal Super-Resolution}

\author{Aditya Kasliwal\textsuperscript{1,4}, Ishaan Gakhar\textsuperscript{2,4}, Aryan Kamani\textsuperscript{1,4}, Pratinav Seth\textsuperscript{1,4}, Ujjwal Verma\textsuperscript{3*}}
\affil{Department of Data Science and Computer Application, Manipal Institute of Technology, Manipal Academy of Higher Education, Manipal, India}
\affil{Department of Information and Communication Technology, Manipal Institute of Technology, Manipal Academy of Higher Education, Manipal, India}
\affil{Department of Electronics and Communication Engineering, Manipal Institute of Technology Bengaluru, Manipal Academy of Higher Education, Manipal, India}
\affil{These authors contributed equally to this work.}
 \corresp{CORRESPONDING AUTHOR: Ujjwal Verma (e-mail: ujjwal.verma@manipal.edu).}



\begin{abstract}
In the last few years, the fusion of multi-modal data has been widely studied for various applications such as robotics, gesture recognition, and autonomous navigation. Indeed, high-quality visual sensors are expensive, and consumer-grade sensors produce low-resolution images. Researchers have developed methods to combine RGB color images with non-visual data, such as thermal, to overcome this limitation to improve resolution. Fusing multiple modalities to produce visually appealing, high-resolution images often requires dense models with millions of parameters and a heavy computational load, which is commonly attributed to the intricate architecture of the model. 

We propose LapGSR, a multimodal, lightweight, generative model incorporating Laplacian image pyramids for guided thermal super-resolution. This approach uses a Laplacian Pyramid on RGB color images to extract vital edge information, which is then used to bypass heavy feature map computation in the higher layers of the model in tandem with a combined pixel and adversarial loss.  LapGSR preserves the spatial and structural details of the image while also being efficient and compact. This results in a model with significantly fewer parameters than other SOTA models while demonstrating excellent results on two cross-domain datasets viz. ULB17-VT and VGTSR datasets.
\end{abstract}


\maketitle

\section{Introduction}
Super-resolution methods have applications in various domains, including digital imagery, photography, robotics, autonomous navigation, surveillance, and security \cite{okarma2015application, nasrollahi2014super, Mei_2021_CVPR}. These techniques aim to enhance the resolution of low-quality images, enabling the extraction of finer details and improved visual quality from lower-resolution source data. Recently, non-visual imagery sensors have gained prominence in various applications due to their ability to capture information beyond the visible spectrum. These sensors, such as thermal cameras, provide crucial data in challenging environments, allowing the detection of heat signatures and other phenomena not perceptible to the human eye. However, thermal images typically have low resolution, and acquiring high-resolution non-visual sensors is expensive. To address this challenge, researchers have delved into Guided Thermal Super Resolution, a field focused on enhancing the resolution of thermal images using RGB images as a guide. With the advent of deep learning, guided super-resolution methods have become increasingly effective, reducing the reliance on expensive, high-quality sensors.

\begin{figure*}[t]
  \centering
  \includegraphics[width=\textwidth]{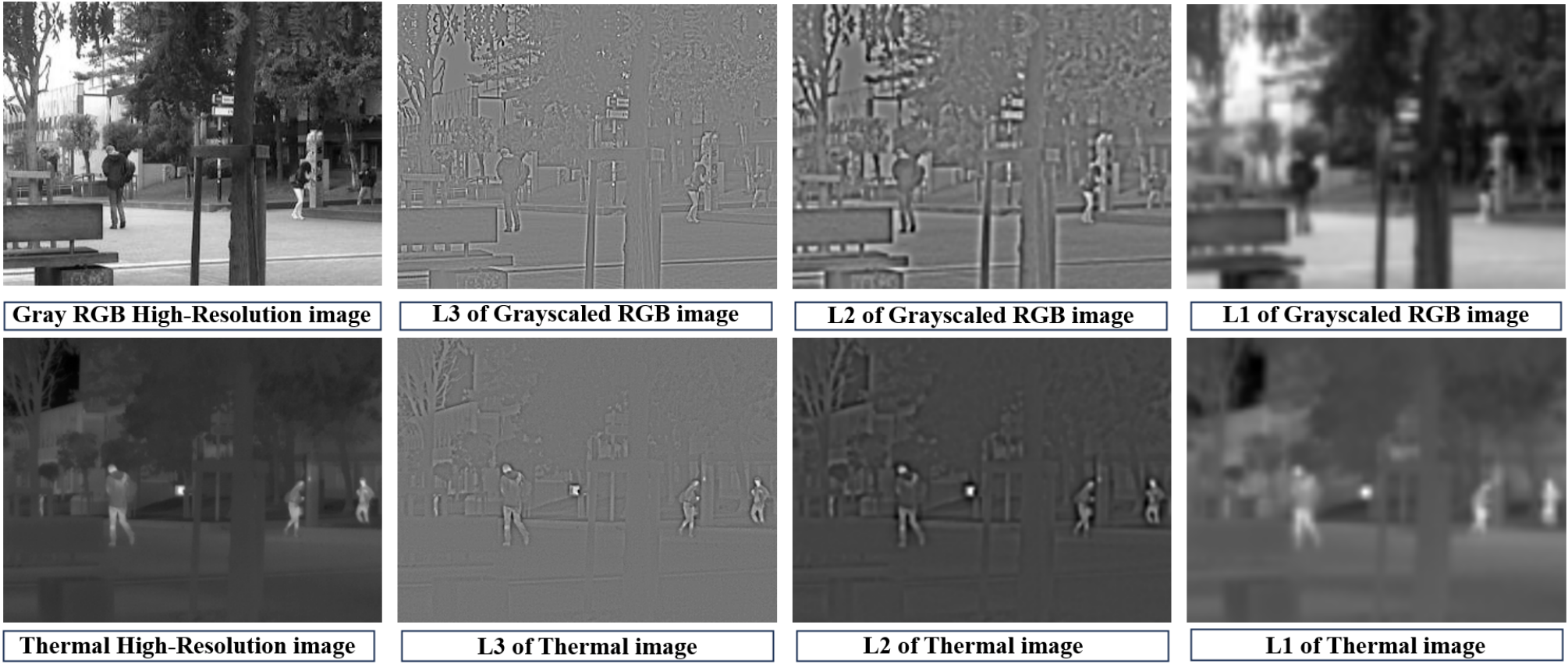}
  \caption{Laplacian Pyramid Visualization. The first row is the Laplacian pyramid of a grayscaled RGB image with two levels and the residual at the end. The second row contains the Laplacian pyramid of a thermal image with two levels and a residual at the end. These images have been taken from the ULB17-VT dataset \cite{almasri2019multimodal}.}
  \label{fig:example}
\end{figure*}

Integrating data from multiple sensors, such as RGB and thermal, can gain more comprehensive insights and enhance performance. The task of enhancing resolution in non-visual data while preserving content remains a persistent challenge \cite{guo2023survey}. 
We experiment with cross-domain datasets, namely the ULB17-VT \cite{almasri2019multimodal} dataset, which is captured by a handheld camera, and the VGTSR dataset \cite{zhao2023thermal}, captured by a UAV (Unmanned Aerial Vision) platform.

To address these issues with a streamlined approach, we introduce LapGSR. LapGSR addresses the challenges of enhancing the resolution of low-quality thermal images by leveraging information from high-resolution RGB images. Our model combines traditional computer vision techniques like Laplacian pyramids with modern deep learning methods for feature extraction. Our framework introduces key components for handling texture, illumination, and other critical features extracted from RGB inputs, ultimately enhancing the resolution of low-quality thermal images. Its architecture incorporates a lightweight, generative neural network with cascading residual blocks, providing robustness against misalignment while ensuring computational efficiency.

Our inspiration for employing Laplacian pyramids stems from their effectiveness in capturing similarities in edge maps between RGB and high-resolution thermal images. As visualized in Figure \ref{fig:example}, the structural features, including texture, intensity, and spatial patterns, are notably congruent within the Laplacian pyramid layers. This observation motivates our use of Laplacian pyramids for feature extraction. We utilize the Laplacian pyramid of our guiding image (RGB) to reconstruct the Laplacian pyramid of the target (thermal) image. We draw upon the Laplacian pyramid technique introduced in prior work \cite{Liang2021HighResolutionPI}, initially devised for addressing image-to-image translation tasks encompassing variations in seasons and lighting conditions. Our proposed model allows for Guided Thermal Super Resolution in scenarios involving alignment and misalignment via the employment of residual blocks in the High Transformation Branch. Our proposed model carries out major feature map transformation in higher layers compared to their model's lowest layers. Through our novel architecture, we achieve adaptive refinement of image structures, resulting in superior image quality. Our contributions extend across various domains, as our model preserves fine details in generated images and has fewer trainable parameters than other SOTA models, making it suitable for real-time applications where image quality is critical. Moreover, the model's adaptability to misaligned thermal images is a significant advantage, ensuring its effectiveness in diverse scenarios. 
Our approach combines traditional computer vision with modern deep learning techniques, achieving state-of-the-art outcomes with fewer parameters compared to recent models in both aligned and misaligned scenarios. 
This highlights its effectiveness and potential influence in the super-resolution domain. 
The main contributions of our proposed method are:
\begin{itemize}
    \item Incorporating the Laplacian image pyramid, a classical computer vision technique, alongside contemporary deep learning approaches for feature extraction, our contribution effectively restores high-resolution images with high visual fidelity and accuracy. This approach allows us to reduce the reliance on a large number of convolutional layers.
    \item A lightweight model architecture that delivers consistent performance on the VGTSR dataset and state-of-the-art performance on the ULB17-VT dataset, with 90\% and 50\% fewer trainable parameters than the SOTA models.
    \item A Multi-Objective Loss that incorporates adversarial loss to address and mitigate the trade-off between PSNR and SSIM
    \item Robustness against misaligned RGB-Thermal pairs, a situation often encountered in real-world scenarios, making our model suitable for deployment.
\end{itemize}

\section{Related Works}
Super Resolution finds applications in security systems, robotic vision, self-driving algorithms, etc. Existing approaches include dictionary-based methods \cite{yang2010image,zeyde2012single} and learning-based methods. A subtask of Super Resolution is Guided Thermal Super Resolution, which involves using a High-Resolution RGB image as a 'guide' for super-resolving the low-resolution thermal image. Thermal Super Resolution finds applications in security and surveillance systems, agriculture, autonomous navigation, etc.

\subsection{CNNs for Super Resolution}
Convolutional Neural Networks (CNNs) for Super Resolution have become increasingly prevalent over the past decade. Their ability to learn complex relationships and hierarchical features, which assist in end-to-end mapping, is the primary reason for their superiority over previous methods. CNNs for Super Resolution were introduced through SRCNN \cite{Dong2014ImageSU}, which learned end-to-end mapping using convolutional layers and achieved superior performance by directly predicting high-resolution patches from low-resolution inputs. VDSR \cite{Kim2015AccurateIS} pioneered a very deep network for Super Resolution. VDSR's novelty lies in its deep architecture with residual connections, effectively mitigating the vanishing gradient problem and enhancing learning at the cost of heavy computation. Using a generator-discriminator setup, SRGAN produced images with improved perceptual quality. Enhanced Deep Super-Resolution (EDSR) \cite{Lim2017EnhancedDR} emphasized depth and width in network design. Its architecture optimization achieved SOTA performance with fewer parameters. However, the robustness of their model is limited since the results are only based on the validation set. ESRGAN \cite{Wang2018ESRGANES} set new benchmarks in visual fidelity, which incorporated perceptual loss, adversarial loss, and a new architecture to enhance SR quality and realism. However, ESRGAN's complex architecture includes dense layers, contributing to its high computational demand during the training and inference phases. The model's ability to generate highly realistic and detailed images comes at the cost of increased computational resources.
The use of GANs in image processing has increased over the past decade due to their ability to generate hyper-realistic images. Their implementation in super-resolution methods has improved the perceptual quality manifold. Laplacian pyramids and transpose convolution for upscaling the coarse-resolution feature maps were first used by \cite{Lai2017DeepLP}. State-of-the-art results on Guided Depth Super Resolution were achieved by \cite{metzger2023guided} by combining guided anisotropic diffusion with a deep convolutional network. 
\subsection{Laplacian-Pyramid-based Super Resolution}
The layers of Laplacian Image Pyramids have been used for super-resolution due to the ability of lower layers to preserve edges and the overall structural integrity of the original image while still ensuring a lightweight model with fast inference times. \cite{Liang2021HighResolutionPI} used a similar Laplacian Pyramid network for image-to-image translation from day-to-night or summer-to-winter. LapSRN \cite{Lai2017DeepLP} used the layers to progressively reconstruct the sub-band residuals of high-resolution images at multiple pyramid levels. However, the reconstruction process in LapSRN involves transpose convolutions, which introduce extra parameters for the reconstruction module. Sharing parameters at every level using the same network \cite{he2018hyperspectral} allowed the parameter number to be independent of the upsampling scale; hence, one single set of parameters is required to construct the network with multiple pyramid levels.

\subsection{Guided Thermal Super Resolution}

Guided Thermal Super Resolution (GTSR) has been gaining relevance due to the many feasible methods proposed in recent years. High-resolution capturing RGB sensors are more economical as compared to IR sensors, and the fusion of these modalities yields a better qualitative and quantitative result than single image thermal super-resolution \cite{Almasri2018MultimodalSF}. To increase the robustness of GTSR models, \cite{Gupta2021TowardUG} incorporated a feature-alignment loss along with a misalignment map alignment block and achieved SOTA results for GTSR on unaligned datasets. Multiple U-Net backbones were used in CoReFusion \cite{Kasliwal_2023_CVPR} to merge multiple modalities while still demonstrating comparable results for cases of missing modalities. However, the model utilizes 20M trainable parameters, owing to the two ResNet-34 encoders used for feature extraction. Self-supervised learning using a transformer to produce SOTA results without artefacts was introduced by \cite{Shacht2020SinglePC}. Using 2 subnetworks for feature extraction from Laplacian pyramids and an attention fusion module, \cite{gupta2020pyramidal} set new qualitative and quantitative results benchmarks. Super-resolution for remote sensing applications where the data from satellite images is fused to generate HR images was also explored by \cite{dong2022learning}. They present findings across various subdomains but do not incorporate any datasets we have utilized. Their work enhances depth, DEM, and thermal imagery resolution. Of these, only the depth imagery originates from a handheld device (Middlebury and NYUv2 datasets), with the remaining datasets acquired through satellite remote sensing. In contrast, our study utilizes datasets from both handheld devices and UAVs for thermal imaging.

\section{Methodology}
We propose LapGSR, a lightweight, generative, robust, multimodal architecture utilizing Laplacian Pyramids for Guided Thermal Super Resolution. Our lightweight model faithfully reconstructs thermal images by the explicit edge guidance provided by the Laplacian decomposition of the RGB image (Fig. \ref{fig:example}).
As depicted in Fig. \ref{fig:MODEL_DIAG}, we decompose the RGB image into a Laplacian pyramid of 3 levels (L3, L2, L1) and replace the last layer (residual, L1) with the low-resolution thermal image ($I_L$). LapGSR employs 3 branches responsible for feature extraction and translation from each layer of the Laplacian pyramid, namely the Lower Transformation Branch, Middle Transformation Branch, and Higher Transformation Branch. Our model suits the SR factor of only 4x. To our knowledge, the GSR domain requires SR of 4x and 8x, and super-resolving by 8x would require another branch.
\begin{figure*}[t]
  \scalebox{0.99}{\centerline{\includegraphics[width=\textwidth, height=4in]{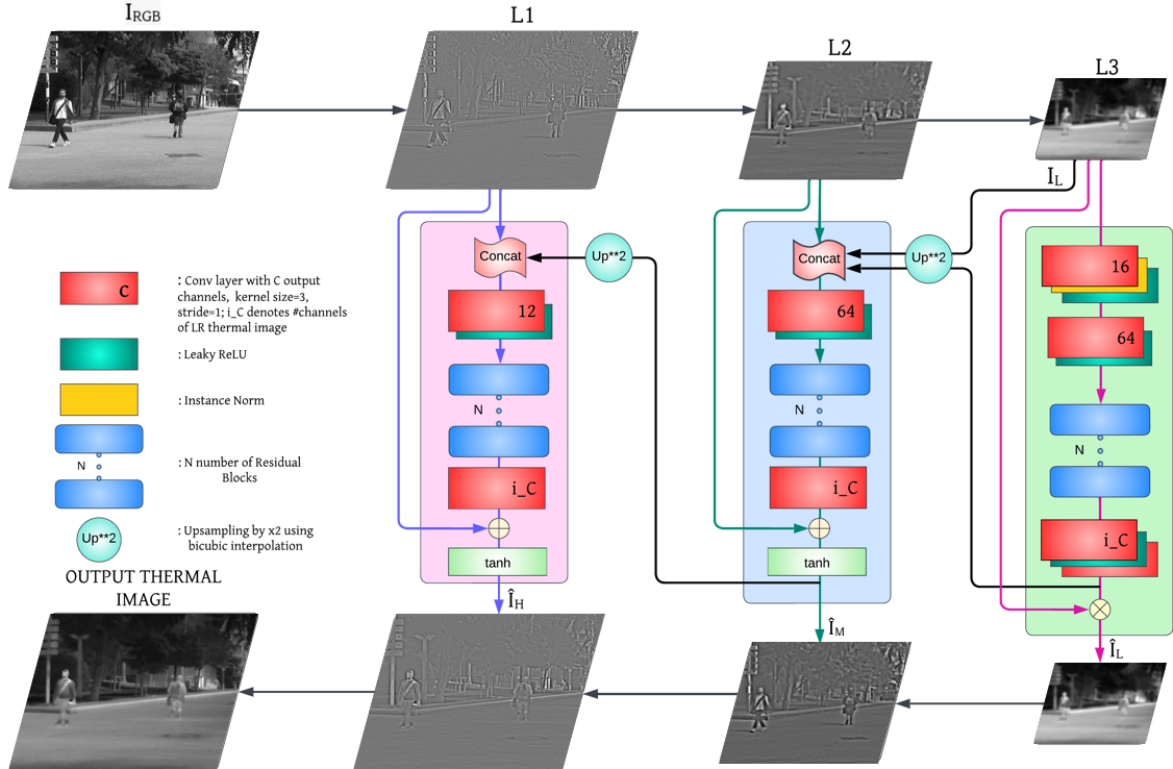}}}
  \caption{Proposed LapGSR model architecture. $I_{RGB}$ refers to the grayscaled high-resolution RGB image from the ULB17-VT dataset. The Green highlight represents the LTB, the Blue represents the MTB, and the Pink represents the HTB. The given figure is for Pyramid of Depth 2; L1, L2, and L3 represent the layers of the modified Laplacian pyramid, where the lower-resolution thermal image replaces the residual.
  The affine transformation is only present in the schematic representation for better visualization and has not been applied to the input or output of the model.}
  \label{fig:lap_vis}
\end{figure*}

\subsection{Lower Transformation Branch}
The Lower Transformation Branch (LTB) distils fundamental features such as luminance, texture, and illumination from the low-resolution thermal image. The LTB consists of a convolutional layer, followed by instance normalization
and a leaky ReLU activation function is applied to prevent the vanishing gradient problem often encountered in deep networks.

After the initial feature extraction, the LTB employs an additional convolutional layer and leaky ReLU activation, followed by a series of residual blocks comprising a convolutional layer, leaky ReLU, another similar convolutional layer, and a skip connection. 

The culmination of the LTB's processing is the generation of a feature map, denoted as \(\hat{I}_L\), which is the lowest layer of the translated pyramid and is also upsampled and combined with the second last layer (L2) of the branch. This enriched feature map is subsequently fed into the Middle Transformation Branch. Finally, the output of the residual blocks is multiplied with L1 to yield the final product of the LTB. By integrating the low-level details extracted by the LTB with the higher-level features, the network can construct a more comprehensive feature representation, setting the stage for advanced processing in subsequent network stages.

\subsection{Middle Transformation Branch}
The Middle Transformation Branch (MTB) serves as a critical component in our network, bridging the low-level feature processing conducted by the Lower Transformation Branch (LTB) and the high-level abstractions required for subsequent tasks. The MTB is tasked with refining and transforming the concatenated feature maps received from the LTB. It also doubles as a realignment module, further demonstrating the robustness of our model against misaligned images. The accentuated edge information and fine details derived from Laplacian pyramids (L2) help to bypass heavy feature map computation. These derived details allow the model to avoid an excessive number of parameters.

The MTB starts with a convolutional layer to extract features accurately without altering the input map's spatial dimensions. Following this layer is a leaky ReLU activation function, which adds non-linearity to the model and mitigates the issue of vanishing gradients. Given that our model has a more refined architecture with fewer convolutional layers than other advanced models, we utilize leaky ReLU to prevent the vanishing gradient problem in our streamlined model. The MTB consist of several residual blocks. A final convolutional layer follows the residual blocks, further refining the feature maps for the activation phase.

The fusion of the low-level (\(\hat{I}_L\)) and intermediate-level features (L2) is then passed through a tanh activation function. 
The output from the tanh activation (\(\hat{I}_M\)) encapsulates a balanced and normalized feature representation, setting the stage for the final transformation tasks within the network's pipeline. The tanh activation is only applied at the end of MTB and HTB since major transformations primarily occur in these branches. The LTB is responsible for extracting minimal features necessary for combining with upper branches and hence has no non-linearity activation.

The middle transformation branch is responsible for extracting spatial and structural information by processing and refining the joint feature maps of the LTB and the L2 layer of the pyramid. This advanced feature representation is crucial for the network's subsequent stages, where such high-level conceptualizations of the image are vital for advanced visual tasks. This contributes to the faithful and perceptually appealing reconstruction of the final image. The output of the MTB (\(\hat{I}_M\)) acts as the intermediate layer of the translated pyramids. It is 2x upsampled using bicubic interpolation and concatenated with L3, where it is subsequently passed into the High Transformation Branch.
\subsection{High Transformation Branch}

The High Transformation Branch (HTB) represents the final stage of our network's feature transformation hierarchy, where the focus shifts to synthesizing the final high-resolution image. This branch is architecturally similar to the Middle Transformation Branch (MTB) but is streamlined to contain lesser residual blocks, reflecting its specialized role in fine-tuning the feature maps for high-resolution output (\(\hat{I}_H\)).

The HTB receives the 2x upsampled output from the MTB (\(\hat{I}_M\)), which carries a rich blend of structural and spatial information. The upsampled feature map enters a convolutional layer that meticulously processes the enlarged feature set. Following this, a leaky ReLU activation function is applied.

The residual blocks at the core of the HTB are critical for refining the upsampled features. The reduced number of blocks compared to the MTB reflects the HTB's role in applying the finishing touches to the image rather than transforming the feature set extensively. 
The 1st convolutional layer of the HTB has a 12-channel output to avoid the high computational complexity in the form of more GFLOPs, which arises as a result of the HTB having to operate on high-resolution features.

After the residual blocks, the feature map undergoes a final convolutional layer, consolidating the high-level features into a coherent image structure. This final convolutional layer is pivotal in rendering the fine details and ensuring that the synthesized high-resolution image retains the textural and structural integrity of the preceding branches. This output is added to L1, and a tanh activation is applied.

The output of the HTB (\(\hat{I}_H\)) is the generation of the top layer of the translated pyramid. This image embodies the detailed textural information from the LTB and integrates the abstract structural and spatial features refined through the MTB, resulting in a comprehensive and high-fidelity visual output.   
\subsection{Inverse Laplacian Reconstruction Module}
To keep our model lightweight and applicable to real-time problems, we use bicubic interpolation to reconstruct the final image from the translated pyramid. The layers (\(\hat{I}_L\)), (\(\hat{I}_M\)), and (\(\hat{I}_H\)) are upsampled and added, in a cascading, inverse-laplacian operation, to reconstruct the final high-resolution thermal image in a non-parametric manner. The final reconstructed image is treated as our model's prediction (\(\hat{y}\)) for the computation of the loss. This inverse operation eliminates transpose convolutions, thereby avoiding the need for more parameters. 
\subsection{Loss Function}
\label{Loss_section}
To ensure the ideal qualitative and quantitative quality of the final high-resolution thermal image (\(\hat{I}_H\)), we introduce a combined loss function to combat the tradeoff between PSNR and SSIM. Our loss function involves pixel-wise $\lambda$-weighted Mean Squared Error Loss (MSELoss) and Adversarial Loss.

\textbf{Mean Squared Error Loss} or MSELoss \cite{goodfellow2016deep} is the pixel-wise difference between the squares of the ground truth and the prediction, divided by the number of samples. This minimization allows for pixel-wise similarity between the ground truth and prediction.

\textbf{Adversarial Loss}
is used to retain visual fidelity through an edge-guided network, we employ adversarial loss or GANLoss. Specifically, we employ a Least-Square Generative Adversarial Network (LSGAN) \cite{Mao2016LeastSG} along with pixel-wise MSE to faithfully reconstruct fine structural and spatial details like illumination, texture, contrast, etc. The output of our model acts as a generator and tries to 'fool' the discriminator. The equations for the Discriminator and Generator are given below:
\begin{equation}
L_{G} = \frac{1}{2} \mathbb{E}_{z \sim p(z)} \left[(D(G(z)) - 1)^2\right]
\end{equation}
Where \( G \) is the generator, \( D \) is the discriminator, \( z \) is a random noise vector sampled in a Gaussian distribution, and \( D(G(z)) \) is the discriminator's prediction for the fake data \cite{goodfellow2014generative} produced by the generator.
\begin{equation}
L_{D} = \frac{1}{2} \mathbb{E}_{x \sim p_{\text{data}}(x)} \left[(D(x) - 1)^2\right] + \frac{1}{2} \mathbb{E}_{z \sim p(z)} \left[D(G(z))^2\right]
\end{equation}
Where \( x \) is real data \cite{goodfellow2014generative}, and \( D(x) \) is the discriminator's prediction for the real data.
The \textbf{Final Loss} involves the weighted addition of MSELoss (reconstruction loss) and GANLoss (adversarial loss) with a balancing hyperparameter $\lambda$. This final loss ensures perceptual quality as well as spatial reconstruction. Hence, the final formula of the loss used in the model is:
\begin{equation}
L = \lambda L_{\text{MSELoss}} + L_{\text{GANLoss}}
\end{equation}

\section{Experiments}

We conduct experiments for high-resolution thermal image generation on two datasets, namely ULB17-VT\cite{almasri2019multimodal} and VGTSR \cite{zhao2023thermal}. While VGTSR has slight misalignment between the RGB-Thermal pair, we also introduce external misalignment on various scales in the ULB17-VT dataset to additionally evaluate our model's robustness.
\subsection{Datasets}
\textbf{ULB17-VT} \cite{almasri2019multimodal} dataset contains mostly well-aligned, low-resolution thermal, high-resolution RGB, and high-resolution thermal images. The authors used a FLIR-E60 camera to capture thermal-visual images with a resolution of 320 x 240 pixels, a thermal sensitivity of 0.05°C, and a range of -20 °C to  650°C. The published benchmark has 404 pairs of images divided into 280 training samples, 78 validation samples, and 46 testing samples. Since the low-resolution thermal image is 1 channel, the input RGB image is grayscaled. It contains scenes of various environments, such as indoor, outdoor, winter, and summer, of static and moving objects. We also experimented by introducing misalignment between the RGB and high-resolution thermal images. Table \ref{Shift Scale Table} demonstrates our results at different scales of misalignment.

\textbf{VGTSR} \cite{zhao2023thermal}, the dataset consists of 1025 pairs of visible, high-resolution, and low-resolution thermal images captured under UAV (Unmanned Aerial Vision) platforms. Each image has a resolution of 640 × 512 and contains a great number of small objects which are manually aligned. The thermal images have a resolution of 160 x 128 and have 3 channels. Hence, the input RGB image is considered a 3-channel image. The visible and thermal UAV images are taken by a DJI Matrice M300 RTK UAV equipped with the Zenmuse H20T sensor, which integrates multiple sensors to capture visible and thermal image pairs simultaneously. The authors of the dataset created the low-resolution images by bicubic and bilinear interpolation. To maintain synonymity with the ULB17-VT, we use the low-resolution thermal images created by bicubic interpolation.

Our choice of datasets further emphasizes the real-world application of our model. The ULB17-VT dataset is captured by a handheld camera, whereas the VGTSR dataset is captured by a UAV (Unmanned Aerial Vision) platform. Our model's ability to maintain results in this cross-domain scenario allows for applicability in actual scenarios.

\subsection{Training Setup}

Through exhaustive experimentation, we decided upon our model's best and most efficient configuration. As mentioned in Section \ref{Loss_section}, our combined loss consisted of a $\lambda$-scaled MSELoss and GANLoss. Given in Table \ref{ULB-LW} are our results of varying values of $\lambda$. The low-resolution thermal and high-resolution RGB images are normalized to a range of [0, 1]. Augmentation is performed on the dataset with vertical and horizontal flipping with a probability of 0.5. We have patched our images, like the authors of the dataset \cite{zhao2023thermal}, to alleviate computational load to a size of 40 x 30. 

Another critical component of our model is the configuration of residual blocks in the LTB, MTB, and HTB. These residual blocks dictate the degree of feature extraction and transformation each branch enables. Hence, we extensively experimented with different numbers of residual blocks in each branch, as displayed in Table \ref{ULB-ResBlock} and Table \ref{UAV-ResBlock}.
Our model utilized the Adam Optimizer \cite{kingma2014adam} and was trained for 100 epochs on the ULB17-VT dataset with a learning rate of $10\textsuperscript{-4}$ and batch size 12 on NVIDIA Tesla P100 GPU, and for 600 epochs on the VGTSR dataset with a learning rate of 3 x $10\textsuperscript{-2}$ on NVIDIA RTX A6000 due the large size of the dataset.

To evaluate our model's results, we use the metrics Peak Signal Noise Ratio (PSNR) \cite{fardo2016formal}, a quantitative measure defined by the ratio of the signal's maximum power of the signal to the power of residual errors and Structural Similarity Index Measure (SSIM) \cite{ndajah2010ssim}, a quantitative measure of spatial reconstruction in image generation. SSIM quantifies the perceptual quality of the generated image.

\section{Results}

\begin{table}[t]
    \caption{Comparison of various GTSR methods against our model on the ULB17-VT dataset. Our model demonstrates state-of-the-art results with significantly fewer parameters. There are spaces left blank because the code for these repositories is not publically available for calculation of parameters.}
    \centering 
    \begin{tabular}{|c|c|c|c|}
    \hline
    \textbf{Method} & \textbf{Params} & \textbf{PSNR} & \textbf{SSIM} \\ \hline \hline
    ZSSR \cite{Shocher2017ZeroShotSU} & - & 26.789 & 0.8567 \\ 
    VTSRGAN \cite{Almasri2018MultimodalSF} & 758K & 27.988 & 0.8202 \\ 
    DeepJF \cite{Li2017JointIF} & - & 27.036 & 0.8410 \\ 
    VTSRCNN \cite{almasri2019multimodal} & 758K & 27.968 & 0.8196 \\
    CMSR \cite{shacht2021single} & - & 29.928 & 0.8820 \\ 
    MMSR \cite{dong2022learning} & - & 30.44 & 0.8984 \\  
    CMPNet \cite{QIAO2023223} &  2.02M & 31.58 & 0.9056 \\ %
    \textbf{LapGSR} & \textbf{398K} &  \textbf{37.516} & \textbf{0.9504} \\ \hline\hline
    \end{tabular}
    \label{ULB RESULTS}
\end{table}

\begin{table}[b]
    \caption{Comparison of state-of-the-art super-resolution methods against our model on the VGTSR dataset.}
    \centering
    \begin{tabular}{|c|c|c|c|}
        \hline
        \textbf{Method}  & \textbf{Params}  & \textbf{PSNR}  & \textbf{SSIM}           \\ \hline \hline
        Restormer \cite{zamir2022restormer} & 25.3M & 30.71 & 0.8933 \\
        MultiNet \cite{han2017convolutional} & 8.7M & 30.31 & 0.8872 \\  
        PAG-SR \cite{gupta2020pyramidal} & 7.6M & 30.74 & 0.8951 \\ 
        UGSR \cite{Gupta2021TowardUG} & 4.5M & 30.29 & 0.8872 \\  
        MGNet \cite{zhao2023thermal} & 18.6M & 31.16 & 0.9033 \\  
        CENet \cite{10401239} & - & 31.24 & 0.9035 \\ %
        \textbf{LapGSR} & \textbf{773K} & \textbf{28.9} & \textbf{0.8878} \\ \hline \hline
        \end{tabular}
        
    \label{UAV-RESULTS}
\end{table}

In this study, we draw comparisons on the ULB17-VT dataset with various guided TSR models. In our experiments with various residual blocks in the lower, middle, and upper transformation branches, we discovered that for the ULB17-VT dataset, a configuration of 2, 3, and 3 residual blocks, respectively, achieves a PSNR of 37.024 and an SSIM of 0.9508 and for the VGTSR dataset, a configuration of 3, 5, and 7 residual blocks, respectively achieves a PSNR of 28.9 and SSIM of 0.887.

\begin{table}[t]
    \caption{Shift scale refers to the fraction by which the RGB image is shifted compared to the high-resolution thermal image. If shift\_limit is a float, the range will be (-shift\_limit, shift\_limit).}
    \centering
    \begin{tabular}{|c|c|c|}
    \hline
        \textbf{Shift Scale} & \textbf{PSNR} & \textbf{SSIM} \\ \hline \hline
        0.05 & 35.174 & 0.9553 \\ 
        0.1 & 36.95 & 0.9559 \\ 
        0.15 & 34.957 & 0.9580\\ 
        \hline \hline
    \end{tabular}
    
    \label{Shift Scale Table}
\end{table}

We present the exceptional outcomes of LapGSR as shown in Table \ref{ULB RESULTS} on the aligned ULB17-VT dataset. Our model not only achieves SOTA results but also remarkably reduces parameter usage to 398K, outlining its efficiency and real-world application.
We compare our model with competing SR methods such as a generative adversarial network (VTSRGAN), a CNN-based network (VTSRCNN) \cite{almasri2019multimodal}, and a joint cross-modality transformer (CMSR) \cite{Shacht2020SinglePC} for the task of guided thermal super-resolution and achieve SOTA results on the ULB17-VT dataset.

\begin{figure*}[tb]{
\centerline
{\includegraphics[width=\textwidth]{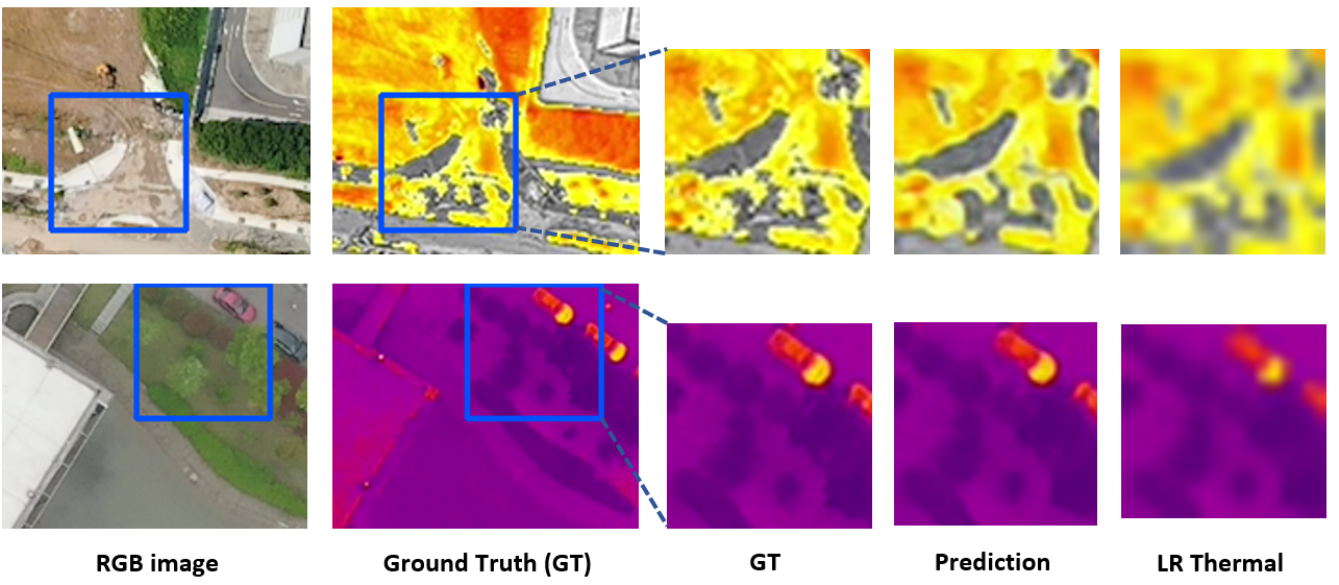}}}
  \caption{visualization of our model's output on 2 instances of the VGTSR dataset. The first image in each row is the high-resolution RGB image, the second is the ground truth, the third is a patch of the Ground Truth, the fourth is a patch of LapGSR's output, and the last is a patch of the low-resolution thermal image.}
  \label{fig:UAV-OUTPUT}
\end{figure*}

Extending our analysis to the VGTSR dataset, our model attains comparable results as shown in Table \ref{UAV-RESULTS}, confirming its prowess in generating high-resolution thermal images. As mentioned by the authors of the dataset in \cite{zhao2023thermal}, UAV images have a larger range of observation due to the shooting height. This leads to UAV images containing more contour information and less texture information (Fig. \ref{fig:UAV-OUTPUT}). Since our model relies on explicit edge-guidance, the lack of textural information leads to slightly diminished performance. 

We establish our model's robustness by evaluating performance on cases of misalignment, primarily on the ULB17-VT dataset, demonstrating our model's ability to maintain consistent super-resolution quality. (Table: \ref{Shift Scale Table}). This robustness is facilitated by augmentations that simulate real-world shifts in perspective. Non-linearities are observed in Table \ref{Shift Scale Table} due to our adopted random shift scaling method. As the caption mentions, random shift scaling shifts every image by any value between (-shift\_limit, shift\_limit). Since there can be a variance in shifting, it may produce a slight variance (5\%) in results.

\

As evident in Fig. \ref{fig:UAV-OUTPUT}, our results are visually apt and similar to the ground truth. The visualization of 2 instances is given row-wise. The first image of each row is the RGB image, the second is the Ground Truth, the third is the patch from the Ground Truth, the fourth is the patch of LapGSR's output, and the last is the low-resolution thermal. On observing the 1st row, it is evident that the RGB image lacks textural information, which results in our model's output being 'smooth' and lacking visual fidelity. The images in the 2nd row contain rich textural information, which results in a highly accurate prediction.

\section{Ablation Study}

In this section, we experimentally study the effects of different parameters and modules on the results of our super-resolution method. We run these experiments on the ULB17-VT and VGTSR datasets. These extensive experiments allow us to pick our model's best configuration of residual blocks and the most balanced loss weight. Although our results on the VGTSR dataset are shown on 600 epochs, we conducted ablations with 400 epochs due to the large computational load.

\begin{table}
\caption{Our results on the ULB17-VT dataset with several configurations of residual blocks. GFPs stands for GigaFLOPs.}
\centering
\begin{tabular}{|c|c|c|c|c|c|c|}
\hline
\multicolumn{3}{|c|}{\textbf{Residual Blocks}} & \multicolumn{1}{c|}{\textbf{Params}} & \multicolumn{1}{|c|}{\textbf{GFPs}} & \multicolumn{1}{c|}{\textbf{Metrics}} \\ 
LTB & MTB & HTB &  &  & PSNR / SSIM \\ \hline \hline
3 & 2 & 3 & 398K & 9.34 & 35.883 / 0.9497  \\
3 & 3 & 3 & 471K & 12.18 & 36.500 / 0.9399  \\
3 & 4 & 3 & 546K & 15.02 & 39.724 / 0.9444  \\
3 & 5 & 3 & 620K & 17.86 & 34.706 / 0.9531  \\
4 & 3 & 3 & 546K & 12.9 & 36.240 / 0.9595  \\
5 & 3 & 3 & 620K & 13.6 & 39.682 / 0.9438  \\
3 & 3 & 2 & 469K & 11.78 & 35.988 / 0.9519  \\
3 & 3 & 4 & 475K & 12.58 & 35.064 / 0.9434  \\
3 & 3 & 5 & 477K & 13 & 39.672 / 0.9449  \\
\textbf{2} & \textbf{3} & \textbf{3} & \textbf{398K} & \textbf{5.74} & \textbf{37.516} / \textbf{0.9504}  \\ \hline \hline
\end{tabular}

\label{ULB-ResBlock}
\end{table}

In Table \ref{ULB-ResBlock}, we present our results on the ULB17-VT dataset with various configurations of residual blocks with $\lambda$ value as 4500.
Through exhaustive experimentation, we observed that the number of residual blocks in the LTB, MTB, and HTB of 2, 3, and 3 demonstrates the best results. A tradeoff exists between the PSNR and SSIM, which is evident when the total number of residual blocks increases, causing either metric to increase at the cost of the other. Moreover, the number of trainable parameters increases beyond the configuration of 2, 3, and 3 while not enhancing performance.

\begin{table}[!htb]
    \caption{Our results on the VGTSR dataset with several configurations of residual blocks. GFPs stands for GigaFLOPs.}
    \centering
        \begin{tabular}{|c|c|c|c|c|c|c|}
            \hline
            \multicolumn{3}{|c|}{\textbf{Residual Blocks}} & \multicolumn{1}{c|}{\textbf{Params}} & \multicolumn{1}{|c|}{\textbf{GFPs}} & \multicolumn{1}{c|}{\textbf{Metrics}} \\ 
            LTB & MTB & HTB &  &  & PSNR / SSIM \\ \hline \hline
            3 & 3 & 3 & 478K & 3.32 & 27.684 / 0.8594\\
            3 & 5 & 3 & 626K & 4.84 & 28.042 / 0.8778\\
            3 & 7 & 3 & 773K & 6.36 & 28.333 / 0.8824\\
            3 & 9 & 3 & 921K & 7.86 & 28.307 / 0.8816\\
            3 & 7 & 1 & 768K & 6.14 & 28.183 / 0.8814 \\
            1 & 7 & 3 & 626K & 5.98 & 28.200 / 0.8779\\
            5 & 7 & 3 & 921K & 6.74 & 28.239 / 0.8827\\
            7 & 7 & 3 & 1.07M & 7.10 & 28.371 / 0.8840\\
            \textbf{3} & \textbf{7} & \textbf{5} & \textbf{779K} & \textbf{6.56} & \textbf{28.354 / 0.8808} \\
            \hline \hline
        \end{tabular}
    \label{UAV-ResBlock}
\end{table}

Directing our attention to the VGTSR dataset, we adopted a similar strategy of exhaustive experimentation for this dataset. We note that the best configuration of the number of residual blocks in the LTB, MTB, and HTB, respectively, are 3, 7, and 5 with $\lambda$ of 4500. An increase in the number of blocks causes a negligible change in performance at the cost of a high number of parameters. As evident in Table \ref{UAV-ResBlock}, the number of residual blocks needed for comparable results is more than for the ULB17-VT dataset. This is because the images in the VGTSR dataset are more complex and have more contour than textural information.

\begin{table}[!ht]
\caption{Our results on the ULB17-VT dataset with various values of $\lambda$.}
\centering
\begin{tabular}{|c|c|c|}
\hline 
\textbf{Loss Weight ($\lambda$)} & \textbf{PSNR} & \textbf{SSIM} \\ \hline \hline
0 & 12.927 & 0.8185 \\
1500 & 36.507 & 0.9491 \\
2500 & 36.033 & 0.9495 \\
3500 & 37.05 &  0.9511 \\
4500 & 37.516 & 0.9504 \\
5500 & 37.418 & 0.9512 \\
6500 & 37.458 & 0.9522 \\ \hline \hline
\end{tabular}
\label{ULB-LW}
\end{table}

As mentioned in Section \ref{Loss_section}, we introduce a combined loss, merging GANLoss and $\lambda$-weighted MSELoss. As evident in Table \ref{ULB-LW}, we conducted thorough experimentation with several values of $\lambda$ on the ULB17-VT dataset. We observe that increasing $\lambda$ to 4500 demonstrates the best results, and increasing it further causes negligible changes in the final PSNR and SSIM. 

In order to obtain the best results, we conducted several experiments with various Adversarial Losses, as mentioned in Table \ref{GAN-Exps}. As evident, the LSGAN demonstrates the best results with the given configuration of residual blocks (2, 3, 3 for ULB17-VT and 3, 7, 5 for VGTSR) and a loss weight of 4500. Hence, we implement the LSGAN Loss in tandem with MSELoss to obtain state-of-the-art results. 


\begin{table}[!ht]
    \centering
    \caption{The results of our model with various GAN loss functions. As mentioned in the results section, the best configurations of Residual Blocks are used in the LTB, MTB and HTB for each dataset, respectively along with the best loss weight.}
        \begin{tabular}{|c|c|c|}
            \hline 
            \textbf{GAN Type} & \textbf{ULB17-VT \cite{almasri2018multimodal}} & \textbf{VGTSR} \cite{zhao2023thermal} \\
            ~ & PSNR / SSIM & PSNR / SSIM \\ \hline \hline
            Vanilla GAN \cite{Goodfellow2021GenerativeAN} & 36.465 / 0.9482 & 28.052 / 0.8803\\
            WGAN \cite{Arjovsky2017WassersteinG} & 36.943 / 0.9489 & 27.814 / 0.8744\\
            Hinge GAN \cite{Lim2017GeometricG} & 36.741 / 0.9486 & 28.167 / 0.8779\\
            \textbf{LSGAN} \cite{Mao2016LeastSG} & \textbf{37.516} / \textbf{0.9504} & \textbf{28.9} / \textbf{0.8878} \\
            \hline \hline
        \end{tabular}
    \label{GAN-Exps}
\end{table}

\section{Qualitative Results}

\begin{figure*}[tb]
    \scalebox{0.90}{
    \centering
    \includegraphics[width=\textwidth, height=4in]{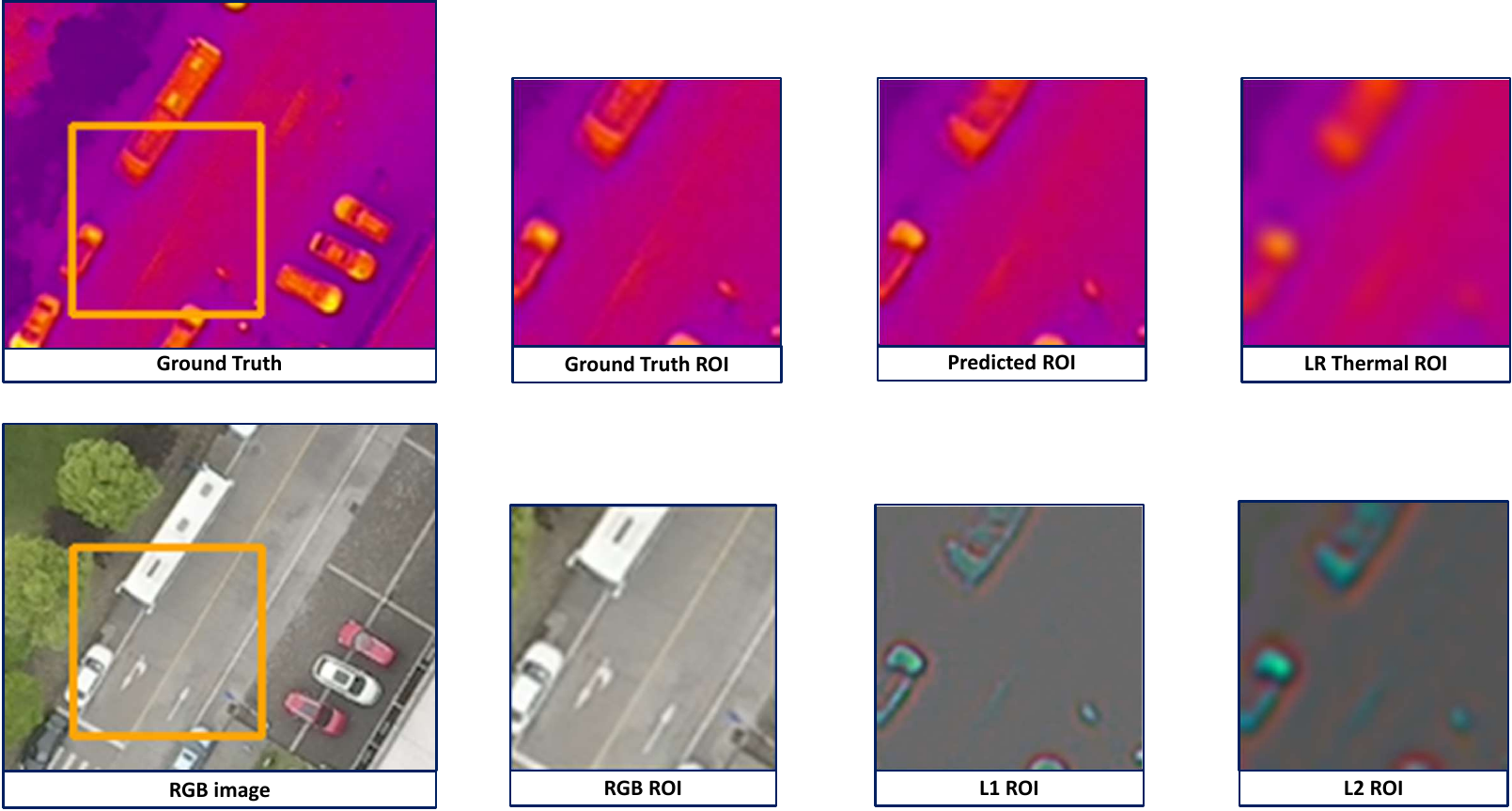}}
    \caption{This is an instance from the VGTSR dataset. The highlighted region of interest (ROI) in the RGB image has sharp edges for the car and bus. This is also present in the ROI of the subsequent Laplacian pyramid, which helps our edge-guided model learn accurate representations of the thermal image, as evident in the predicted ROI. L1 and L2 stand for the first and second layers of the Laplacian Pyramid.}
    \label{fig:P1}
\end{figure*}

\begin{figure*}
    \small
    \scalebox{0.99}{
    \centering
    \includegraphics[width=\textwidth]{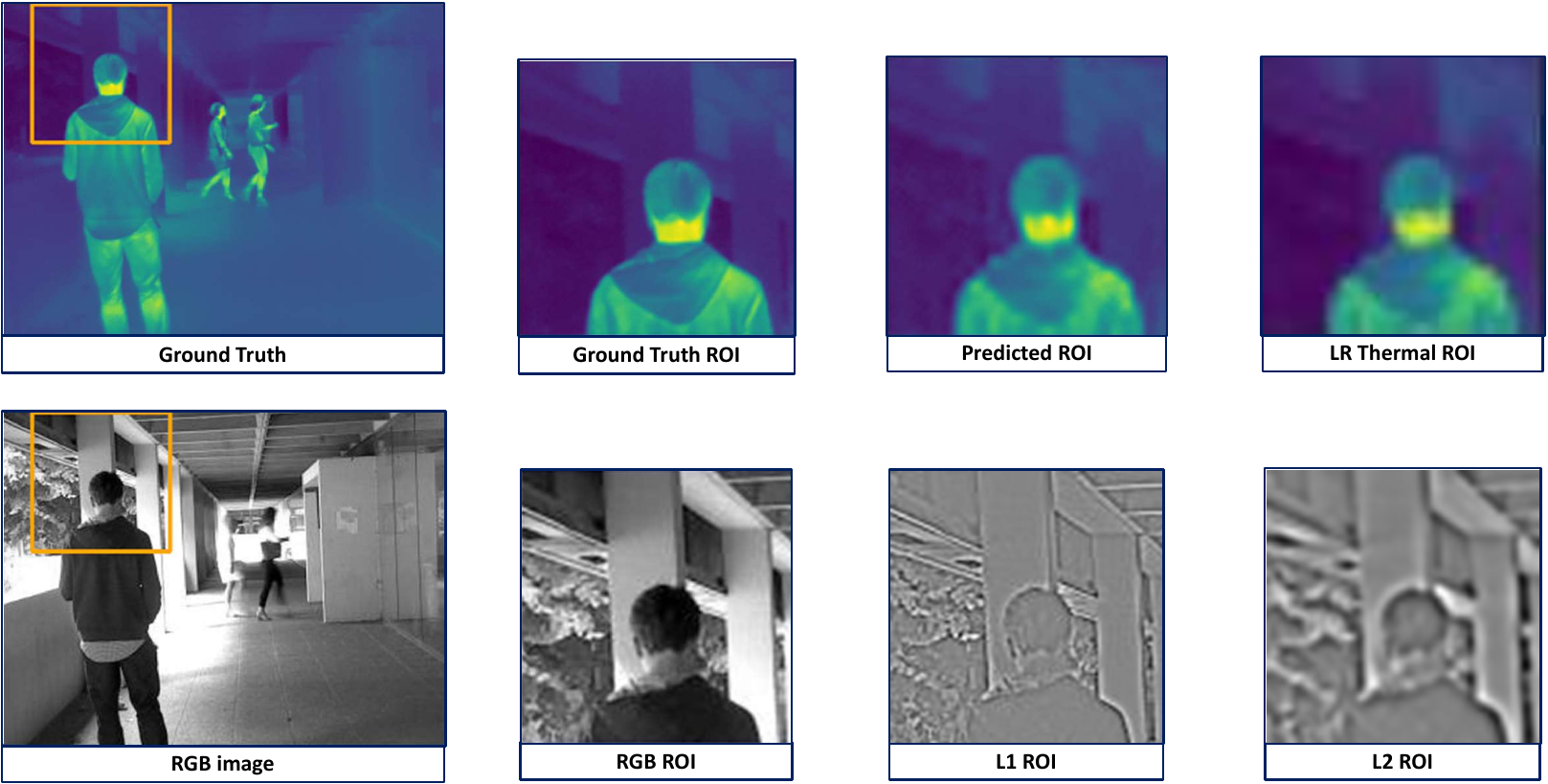}}
    \caption{This is a patch from an instance of the ULB17-VT dataset. The highlighted region-of-interest (ROI) in the RGB image has precise edges of the man. This abundance of textural information is also present in the ROI of the subsequent Laplacian pyramid, which results in excellent results for our model, as evident in the Predicted ROI. L1 and L2 stand for the first and second layers of the Laplacian Pyramid.}
    \label{fig:P1ULB}
\end{figure*}

Our model accurately reconstructs thermal images when the 'guide' RGB image contains adequate textural information, as demonstrated by Fig. \ref{fig:P1ULB} and Fig. \ref{fig:P1}.
For each instance, the low-resolution thermal region of interest has been bilinearly upsampled for clear visualization.

\begin{figure*}
    \small
    \scalebox{0.99}{
    \centering
    \includegraphics[width=\textwidth]{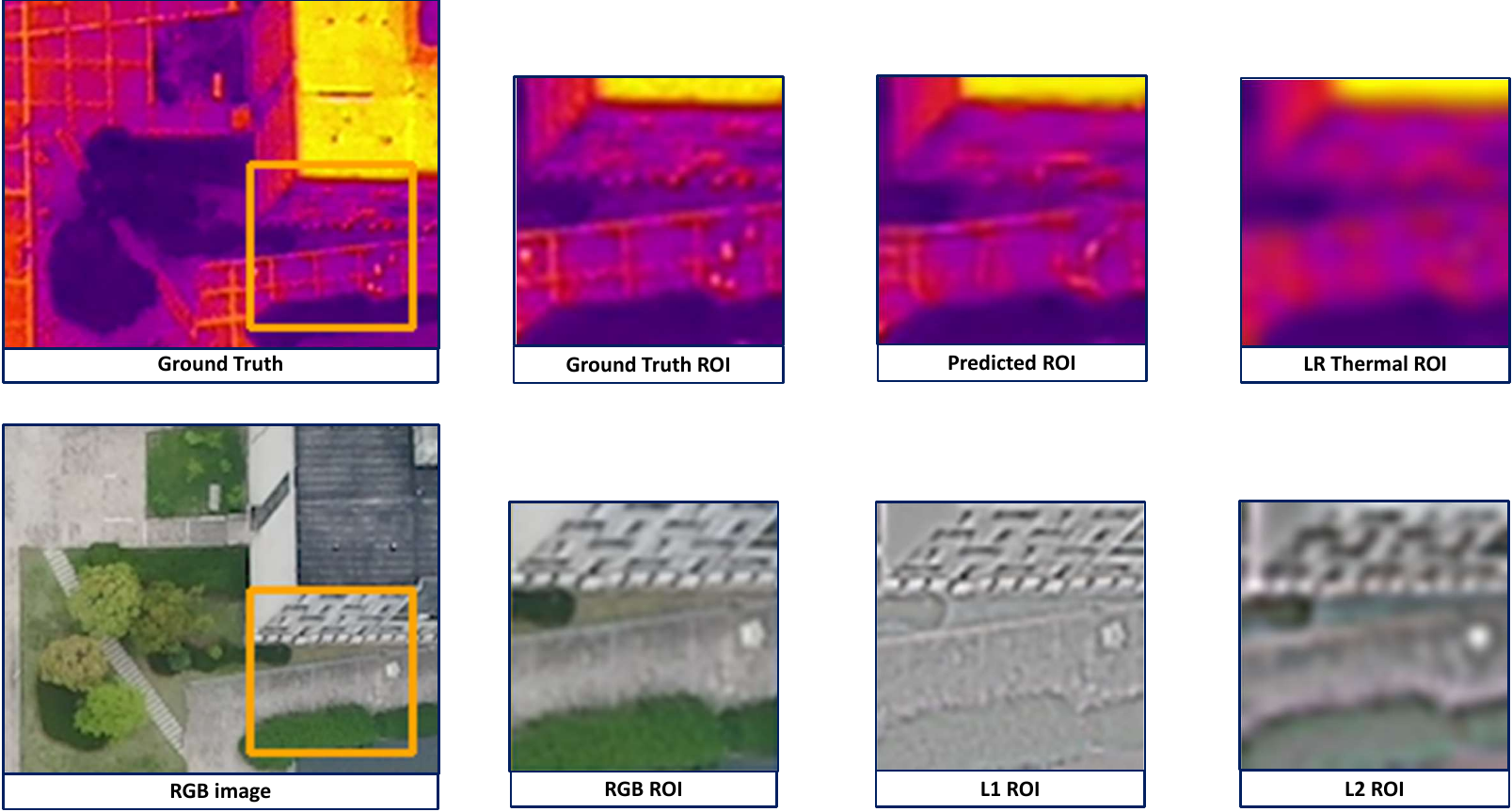}}
    \caption{This is a patch from an instance of the VGTSR dataset. The highlighted region of interest (ROI) in the RGB image has blurry driveway edges. This lack of textural information is also present in the ROI of the subsequent Laplacian pyramid, which results in sub-par results of our model, evident in the Predicted ROI. L1 and L2 stand for the first and second layers of the Laplacian Pyramid.}
    \label{fig:N1}
\end{figure*}

Fig. \ref{fig:P1} is an instance from the VGTSR dataset. In this case, the RGB image contains clear edge information, as depicted in its Laplacian layers, L1 and L2. This allows the model to learn features and accurately reconstruct a high-resolution thermal image. However, Fig. \ref{fig:N1}  contains more contour information and less textural information. 
In Fig. \ref{fig:N1}, an instance of the VGTSR dataset, the exact details of the road/driveway are missing in the RGB image and the L1 and L2 layers of its Laplacian pyramid. This leads to these details being absent in our model's prediction.
Fig. \ref{fig:P1}, an instance of the VGTSR dataset, contains sharp details reflected in the RGB image's Laplacian pyramids, which allows our model to predict accurate and visually apt results.

\begin{figure*}
    \small
    \scalebox{0.99}{
    \centering
    \includegraphics[width=\textwidth]{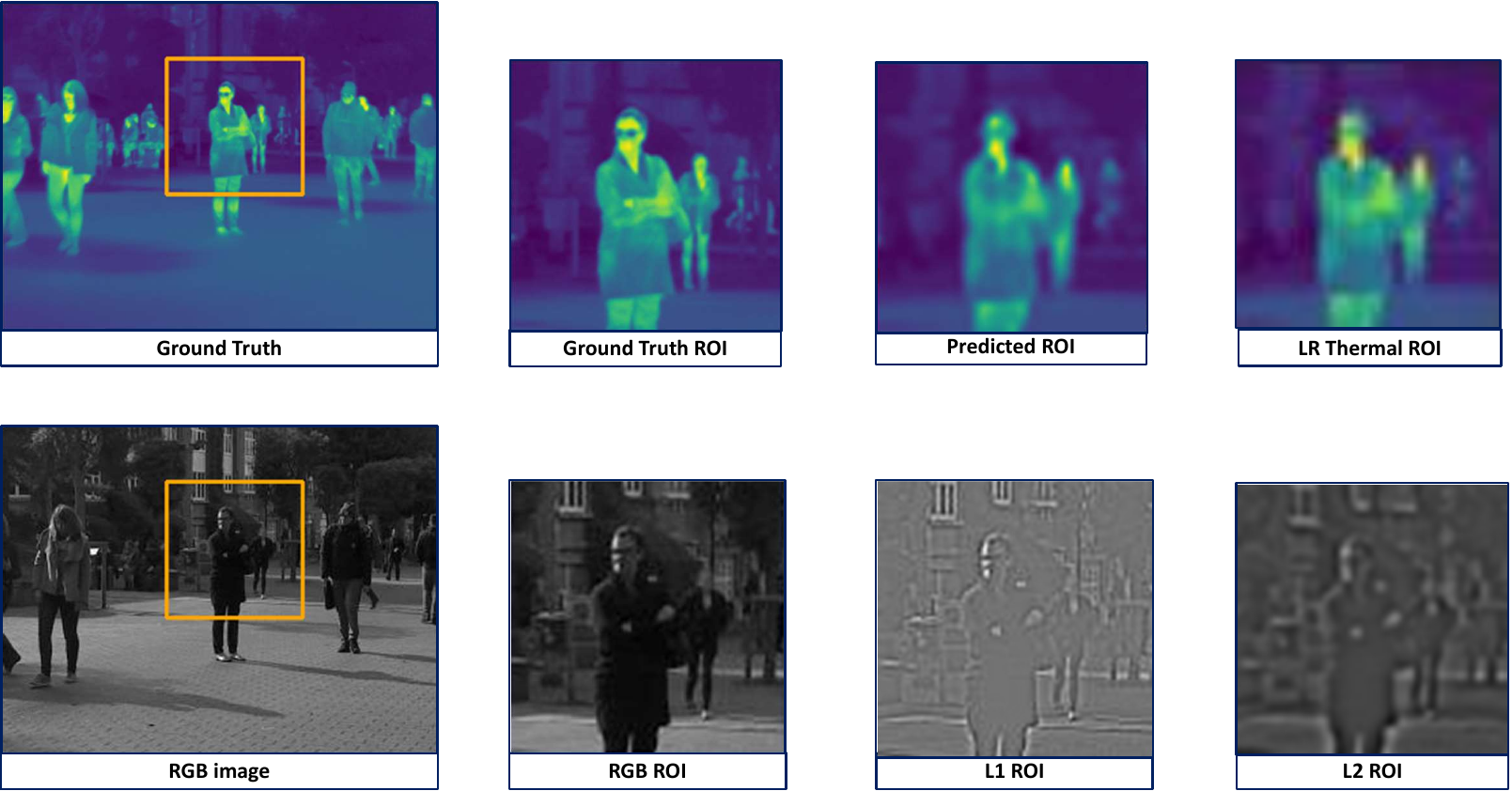}}
    \caption{This is a patch from an instance of the ULB17-VT dataset. The highlighted region of interest (ROI) in the RGB image has blurred edges of the person's arms and spectacles. This consistency of texture results in poor edge information, as visualized in the ROI of the Laplacian Pyramids. Hence, this results in slightly compromised results, as evident in the Predicted ROI. L1 and L2 stand for the first and second layers of the Laplacian Pyramid.}
    \label{fig:N1ULB}
\end{figure*}

Fig. \ref{fig:P1ULB}, taken from the ULB17-VT dataset, contains adequate textural enough to allow our model to capture finer details and reproduce a visually appealing image. The RGB images in both figures contain explicit edge details, evident in the Laplacian pyramid's layers, allowing our model to generate more realistic and accurate outputs.
Fig. \ref{fig:N1ULB}, also taken from the ULB17-VT dataset, demonstrates subpar results. This is because, on observing the RGB image, it is noted that no edges are present on the man's torso in the ROI. This results in its Laplacian pyramid's layers also containing minimal edge information, leading to our model being unable to capture finer details.

Hence, it is evident from all our examples that instances where edge information is subpar lead to only passable results for our model. Although these images are seldom present in the ULB17-VT dataset, they are ubiquitous in the VGTSR dataset.



\section{Conclusion and Future Works}

We propose LapGSR, a lightweight model that uses Laplacian image pyramids to improve the resolution of thermal images. The model decomposes the original RGB image into a modified Laplacian pyramid, which preserves finer image details. This bypasses the need for heavy, high-resolution feature map computation, resulting in a condensed and efficient model. LapGSR can produce SOTA results while being robust and lightweight (398K parameters on the ULB17-VT dataset and 773K on the VGTSR dataset), making it suitable for real-world applications.

One of the limitations of our proposed method is its performance with aerial view images. This is because our model uses a non-parametric Laplacian reconstruction module to generate the final thermal image. This module cannot account for the loss of edge information that occurs in RGB images when the objects are at a distance. In the future, we plan to address this limitation by making provisions for feature extraction from high-altitude RGB images. 
This would allow the architecture to better handle UAV images and produce more accurate super-resolved thermal images. We plan to make our codebase open-source upon the acceptance of our paper.

\section{Acknowledgement}

We thank Mars Rover Manipal, an interdisciplinary student team at Manipal Institute of Technology, Manipal Academy of Higher Education, for the resources needed to conduct our research.

{
    \small
    \bibliographystyle{ieeetr}
    \bibliography{ojim.bib}
}

\begin{IEEEbiography}[{\includegraphics[width=1in,height=1.25in,clip,keepaspectratio]{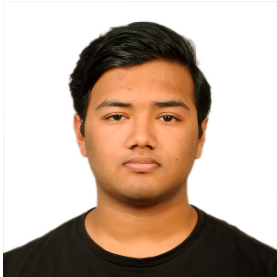}}] {Aditya Kasliwal } is a final-year undergraduate student pursuing  B.Tech in Data Science and Engineering at Manipal Institute of Technology, Karnataka, India. He is a student researcher at Mars Rover Manipal, contributing to AI research. Aditya has interned at organizations like CarDekho and RealtyAI, where he worked on projects involving computer vision and deep learning. His research interests include computer vision, multi-modal deep learning, medical imaging, and federated learning.

\end{IEEEbiography}

\begin{IEEEbiography}[{\includegraphics[width=1in,height=1.25in,clip,keepaspectratio]{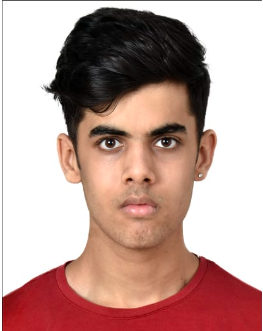}}] {Ishaan Gakhar } is an undergraduate student pursuing B. Tech in Information Technology at Manipal Institute of Technology, Karnataka, India. He is a student researcher at the inter-disciplinary rover team in Manipal, called Mars Rover Manipal. His interests in research lie in Computer Vision and Machine Learning.
\end{IEEEbiography}

\begin{IEEEbiography}[{\includegraphics[width=1in,height=1.25in,clip,keepaspectratio]{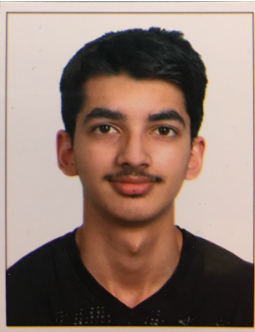}}] {Aryan Kamani } is a fourth-year undergraduate student at Manipal Institute of Technology, pursuing  B.Tech in Data Science. His research interests focus on advanced techniques in computer vision, including Super Resolution, Few Shot Medical Image Classification and Segmentation, and Generative Adversarial Networks (GANs). With a passion for leveraging AI in medical applications, he is committed to exploring innovative solutions in the field of medical imaging.
\end{IEEEbiography}

\begin{IEEEbiography}[{\includegraphics[width=1in,height=1.25in,clip,keepaspectratio]{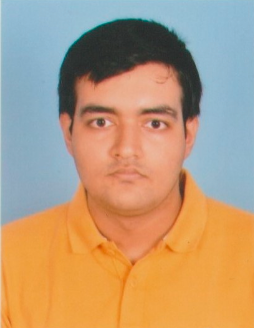}}] {Pratinav Seth } graduated from the Department of Data Science and Computer Applications at Manipal Institute of Technology in 2024. He is currently a research scientist at Arya.ai. From 2021 to 2023, he contributed to the research efforts at Mars Rover Manipal. His research interests include trustworthy machine learning, deep learning for multi-modal applications, medical imaging, geospatial applications for social good, and generative AI, particularly in image generation and natural language processing.

\end{IEEEbiography}

\begin{IEEEbiography}[{\includegraphics[width=1in,height=1.25in,clip,keepaspectratio]{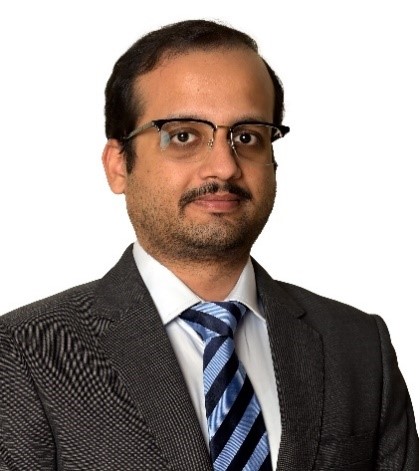}}]{Ujjwal Verma } (Senior Member, IEEE) received his Ph.D. from Télécom ParisTech, University of Paris-Saclay, Paris, France, in Image Analysis and his M.S. (Research) from IMT Atlantique (France) in Signal and Image Processing. Dr. Verma is currently an Associate Professor and Head of the Department of Electronics and Communication Engineering at Manipal Institute of Technology, Bengaluru, India. His research interests include Computer Vision and Machine Learning, focusing on variational methods in image segmentation, deep learning methods for scene understanding, and semantic segmentation of aerial images. He is a recipient of the "ISCA Young Scientist Award 2017-18" by the Indian Science Congress Association (ISCA),  a professional body under the Department of Science and Technology, Government of India. Dr. Verma is the Co-Lead for the Working Group on Machine/Deep Learning for Image Analysis (WG-MIA) of the Image Analysis and Data Fusion Technical Committee (IADF TC) of the IEEE Geoscience and Remote Sensing Society. He is an Associate Editor for IEEE Geoscience and Remote Sensing Letters. He is a Guest Editor for Special Issue in  IEEE Journal of Selected Topics in Applied Earth Observations and Remote Sensing and a reviewer for several journals (IEEE Transactions on Image Processing, IEEE Transactions on Geoscience and Remote Sensing, IEEE Geoscience and Remote Sensing Letters). He is also a Sectional Recorder for the ICT Section of the Indian Science Congress Association for 2020-24. Dr. Verma is a Life Member of the Indian Science Congress Association.
\end{IEEEbiography}

\end{document}